\hoffset=10truemm
\hsize=144truemm
\vsize=230truemm
\pretolerance=5000
\raggedbottom
\baselineskip=12.045truept
\parindent=2.5em
\pageno=0
\def\ls{\vskip 12.045pt}

\font\elevenrm=cmr10 at 11truept
\font\it=cmti10 at 11truept
\font\bf=cmbx10 at 11truept
 at 10 truept
\elevenrm
\ \
\ls\ls\ls
\ls
\ls
\ls
\centerline{\bf A Few Things We Do Not Know About the Sun and F Stars 
and G Stars}
\ls\ls\ls
\centerline{Robert L. Kurucz}
\centerline{Harvard-Smithsonian Center for Astrophysics}
\centerline{60 Garden Street, Cambridge, MA 02138, USA}
\ls
\ls
\centerline{October 3, 1999}
\ls
\ls
\ls\ls\ls
\centerline{presented at the Workshop on Nearby Stars, 
NASA Ames Research Center,}
\centerline{Moffett Field, California, June 24-26, 1999.}
\vfill
\eject

\noindent
{\bf A Few Things We Do Not Know About the Sun and F Stars and G Stars}
\ls
\noindent
Robert L. Kurucz
 
\noindent
Harvard-Smithsonian Center for Astrophysics
 
\noindent
Cambridge, MA 02138, USA
\ls
\ls
\noindent
{\bf We do not know how to make realistic model atmospheres}
 
\noindent
{\bf We do not understand convection}
\ls

Recently I have been preoccupied with convection because the
model atmospheres are now good enough to show shortcomings in the
convective treatment.  Here I will outline what I have learned.  I will
mainly list the conclusions I have come to from examining individual
convective models and from examining grids of convective models as a whole.
Eighteen figures illustrating the points made here can be found in Kurucz 
(1996).
 
     Every observation, measurement, model, and theory has seven 
characteristic numbers: resolution in space, in time, and in energy, and 
minimum and maximum energy.  Many people never think about these resolutions.  
A low resolution physics cannot be used to study something in which the 
physical process of interest occurs at high resolution unless the high 
resolution effects average out when integrated over the resolution bandpasses.
 
     What does the sun, or any convective atmosphere, actually look like?  
We do not really know yet.  There is a very simplified three-dimensional
radiation-hydrodynamics calculation discussed in the review by Chan, Nordlund, 
Steffen, and Stein (1991).  It is consistent with the high spatial and 
temporal resolution observations shown in the review by Topka and Title (1991).  
Qualitatively, there is cellular convection with relatively slowly ascending, 
hot, broad, diverging flows that turn over and merge with their neighbors to 
form cold, rapidly descending, filamentary flows that diffuse at the bottom.  
The filling factor for the cold downward flowing elements is small.  The 
structure changes with time.  Nordlund and Dravins (1990) discuss four 
similar stellar models with many figures.  Every one-dimensional mixing-length 
convective model is based on the assumption that the convective structure 
averages away so that the emergent radiation depends only a one-dimensional 
temperature distribution.
 
     There is a solar flux atlas (Kurucz, Furenlid, Brault, and Testerman 
1984) that Ingemar Furenlid caused to be produced because he wanted to work 
with the sun as a star for comparison to other stars.  The atlas is pieced 
together from eight Fourier transform spectrograph scans, each of which was 
integrated for two hours, so the time resolution is two hours for a given 
scan.  The x and y resolutions are the diameter of the sun.  The z resolution 
(from the formation depths of features in the spectrum) is difficult to 
estimate.  It depends on the signal-to-noise and the number of resolution 
elements.  The first is greater than 3000 and the second is more than one 
million.  It may be possible to find enough weak lines in the wings and 
shoulders of strong lines to map out relative positions to a few kilometers.  
Today I think it is to a few tens of kilometers. The resolving power is on 
the order of 522,000.  This is not really good enough for observations made 
through the atmosphere because it does not resolve the terrestrial lines that 
must be removed from the spectrum.  (In the infrared there are many wavelength 
regions where the terrestrial absorption is too strong to remove.)
The sun itself degrades its own flux spectrum by differential rotation and
macroturbulent motions.  The energy range of the atlas is from 300 to 1300 nm,
essentially the range where the sun radiates most of its energy.
 
     This solar atlas is of higher quality than any stellar spectrum taken 
thus far but still needs considerable improvement.  If we have difficulty 
interpreting these data, it can only be worse for other stars where the 
spectra are of lower quality by orders of magnitude.
 
     To analyze this spectrum, or any other spectrum,  we need a theory that
works at a similar resolution or better.  We use a plane parallel,
one-dimensional theoretical or empirical model atmosphere that extends in z
through the region where the lines and continuum are formed.  The one-dimensional
model atmosphere represents the space average of the convective structure
over the whole stellar disk (taking account of the center-to-limb variation)
and the time average over hours.    It is usually possible to compute a model
that matches the observed energy distribution around the flux maximum.  However,
to obtain the match it is necessary to adjust a number of free parameters:
effective temperature, surface gravity, microturbulent velocity, and the
mixing-length-to-scale-height-ratio in the one-dimensional convective
treatment.  The microturbulent velocity parameter also produces an adjustment
to the line opacity to make up for missing lines.  Since much of the spectrum
is produced near the flux maximum, at depths in the atmosphere where the
overall flux is produced, averaging should give good results.  The parameters
of the fitted model may not be those of the star, but the radiation field
should be like that of the star.  The sun is the only star where the effective
temperature and gravity are accurately known.  In computing the detailed
spectrum, it is possible to adjust the line parameters to match many features,
although not the centers of the strongest lines.  These are affected
by the chromosphere and by NLTE.  Since very few lines have atomic data
known accurately enough to constrain the model,
a match does not necessarily mean that the model is correct.
 
     From plots of the convective flux and velocity for grids of models I
have identified three types of convection in stellar atmospheres:
 
\noindent $\bullet$ normal strong convection where the convection is continuous
from the atmosphere down into the underlying envelope.  Convection carries
more than 90\% of the flux.    Stars with effective temperatures 6000K and
cooler are convective in this way as are stars on the main sequence up to
8000K.  At higher temperature the convection carries less of the total flux
and eventually disappears starting with the lowest gravity models.
Intermediate gravities have intermediate behavior.  Abundances have to be
uniform through the atmosphere into the envelope.  The highly convective
models seem to be reasonable representations of real stars, except for the
shortcomings cited below.
 
\noindent $\bullet$ atmospheric layer convection where, as convection weakens,
the convection zone withdraws
completely up from the envelope into the atmosphere.  There is zero convection
at the bottom of the atmosphere.  Abundances in the atmosphere are decoupled
from abundances in the envelope.  For mixing-length models the convection
zone is limited at the top by the Schwarzschild criterion to the vicinity
of optical depth 1 or 2.  The convection zone is squashed into a thin layer.
In a grid, this layer continues to carry significant convective flux for about
500K in effective temperature beyond the strongly convective models.  There is
no common-sense way in which to have convective motions in a thin layer in an
atmosphere.  The solution is that the Schwarzschild criterion does not apply
to convective atmospheres.  The derivatives are defined only in one dimensional
models.  A real convective element has to decide what to do on the basis of
local three-dimensional derivatives, not on means.
These thin-layer-convective model atmospheres may not be very realistic.
 
\noindent $\bullet$  plume convection.  Once the convective flux drops to the
percent range, cellular convection is no longer viable.  Either the star becomes
completely radiative, or it becomes radiative with convective plumes that cover only
a small fraction of the surface in space and time.  Warm convective material
rises and radiates.  The star has rubeola.   The plumes dissipate and the
whole atmosphere relaxes downward.  There are no downward flows.  The convective
model atmospheres are not very realistic except when the convection is so small
as to have negligible effect, i.e. the model is radiative.  The best approach
may be simply to define a star with less than, say, 1\% convection as radiative.
The error will probably be less than using mixing-length model atmospheres.
 
      Using a one-dimensional model atmosphere to represent a real convective
atmosphere for any property that does not average in space and time to the
one-dimensional model predictions produces systematic errors.  The Planck function,
the Boltzmann factor, and the Saha equation are functions that do not average
between hot and cold convective elements.  We can automatically conclude that
one-dimensional convective models must predict the wrong value for any parameter
that has strong exponential temperature dependence from these functions.
 
     Starting with the Planck function, the ultraviolet photospheric flux in any
convective star must be higher than predicted by a one-dimensional model (Bikmaev 1994).
Then, by flux conservation, the flux redward of the flux maximum must be lower.
It is fit by a model with lower effective temperature than that of the star.
The following qualitative predictions result from the exponential
falloff of the flux blueward of the flux maximum:
 
\noindent $\bullet$ the Balmer continuum in all convective stars is higher than predicted by
a one-dimensional model;
 
\noindent $\bullet$ in G stars, including the sun, the discrepancy reaches up to about 400nm;
 
\noindent $\bullet$ all ultraviolet photoionization rates at photospheric depths are higher
in real stars than computed from one-dimensional models;
 
\noindent $\bullet$ flux from a temperature minimum and a chromospheric temperature rise
masks the increased photospheric flux in the ultraviolet;
 
\noindent $\bullet$ the spectrum predicted from a one-dimensional model for the exponential
falloff region, and abundances derived therefrom, are systematically in error;
 
\noindent $\bullet$ limb-darkening predicted from a one-dimensional model for the exponential
falloff region is systematically in error;
 
\noindent $\bullet$ convective stars produce slightly less infrared flux than
do one-dimensional models.
 
    The Boltzmann factor is extremely temperature sensitive for highly excited levels:
 
\noindent $\bullet$ the strong Boltzmann temperature dependence of the second level of
hydrogen implies that the Balmer line wings are preferentially formed in the
hotter convective elements.  A one-dimensional model that matches Balmer line
wings has a higher effective temperature than the real star;
 
\noindent $\bullet$ the same is true for all infrared hydrogen lines.
 
     The Saha equation is safe only for the dominant species:
 
\noindent $\bullet$ neutral atoms for an element that is mostly ionized are
the most dangerous because (in LTE) they are much more abundant in the cool
convective elements.  When Fe is mostly ionized the metallicity determination
from Fe I can be systematically offset and can result in a systematic error
in the assumed evolutionary track and age.
 
\noindent $\bullet$ in the sun convection may account for the remaining
uncertainties with Fe I found by Blackwell, Lynas-Gray, and Smith (1995);
 
\noindent $\bullet$ the most striking case is the large systematic error in
Li abundance determination in extreme Population II G subdwarfs.  The abundance
is determined from the Li I D lines which are formed at depths in the highly
convective atmosphere where Li is 99.94\% ionized (Kurucz 1995b);
 
\noindent $\bullet$ molecules with high dissociation energies such as CO are
also much more abundant in the cool convective elements.  The CO fundamental
line cores in the solar infrared are deeper than any one-dimensional model
predicts (Ayres and Testerman 1981) because the cooler convective elements
that exist only a short time have more CO than the mean model.
 
     Given all of these difficulties, how should we proceed?  One-dimensional
model atmospheres can never reproduce real convective atmospheres.  The
only practical procedure is to compute grids of model atmospheres, then to
compute diagnostics for temperature, gravity, abundances, etc., and then
to make tables of corrections.  Say, for example, in using the H$\alpha$ wings
as a diagnostic of effective temperature in G stars, the models may predict
effective temperatures that are 100K too high.  So if one uses
an H$\alpha$ temperature scale it has to be corrected by 100K to give the true
answer.  Every temperature scale by any method has to be corrected in some
way.  Unfortunately, not only is this tedious, but it is very difficult or
impossible because no standards exist.  We do not know the energy distribution
or the photospheric spectrum of a single star, even the sun.  We do not know
what spectrum corresponds to a given effective temperature, gravity, or
abundances.  The uncertainties in solar abundances are greater than 10\%,
except for hydrogen, and solar abundances are the best known.  It is crucial
to obtain high resolution, high signal-to-noise observations of the bright
stars.
\ls
\ls
\noindent
{\bf We do not consider the variation in microturbulent velocity}
\ls
Microturbulent velocity in the photosphere is just the convective motions.
At the bottom of the atmosphere it is approximately the maximum convective
velocity.  At the temperature minimum it is zero or near zero because the
convecting material does not rise that high.  There is also microturbulent
velocity in the chromosphere increasing outward from the temperature minimum
that is produced by waves or other heating mechanisms.   In the sun the
empirically determined microturbulent velocity is about 0.5 km/s at the
temperature minimum and about 1.8 km/s in the deepest layers we can see.
In a solar model the maximum convective velocity is 2.3 km/s.  The maximum
convective velocity is about 0.25 km/s in an M dwarf and increases up the
main sequence.  The convective velocity increases greatly as the gravity
decreases.  I suggest that a good way to treat the behavior of microturbulent
velocity in the models is to scale the solar empirical distribution as a
function of Rosseland optical depth to the maximum convective velocity for
each effective temperature and gravity.
 
Why does this matter?  Microturbulent velocity increases line width and
opacity and produces effects on an atmosphere like those from changing
abundances.  At present, models, fluxes, colors, spectra, etc are computed with
constant microturbulent velocity within a model and from model to model.
This introduces systematic errors within a model between high and low
depths of formation, and between models with different effective temperatures,
and between models with different gravity.  Microturbulent velocity varies 
along an evolutionary track.  If microturbulent velocity is
produced by convection, microturbulent velocity is zero when there is no
convection, and diffusion is possible.
 
 By now I should have computed a model grid with varying microturbulent
velocity but I am behind as usual.
\ls
\ls
\noindent
{\bf We do not understand spectroscopy}
 
\noindent
{\bf We do not have good spectra of the sun or any other star}
\ls
Very few of the features called ``lines" in a spectrum are single lines.
Most features consist of blends of many lines from different atoms and
molecules.  All atomic lines except those of thorium have hyperfine or
isotopic components, or both, and are asymmetric (Kurucz 1993).  Low resolution,
low-signal-to-noise spectra do not contain enough information in themselves
to allow interpretation.  Spectra cannot be properly interpreted without
signal-to-noise and resolution high enough to give us all the information
the star is broadcasting about itself.  And then we need laboratory data
and theoretical calculations as complete as possible.
Once we understand high quality spectra we can look at other stars with
lower resolution and signal-to-noise and have a chance to make sense of them.
\ls
\ls
\noindent
{\bf We do not have energy distributions for the sun or any other star}
\ls
I get requests from people who want to know the solar irradiance spectrum,
the spectrum above the atmosphere, that illuminates all solar system bodies.
They want to interpret their space telescope observations or work on
atmospheric chemistry, or whatever.  I say, ``Sorry, it has never been
observed.  NASA and ESA are not interested.  I can give you my model
predictions but you cannot trust them in detail, only in, say, one wavenumber
bins."  The situation is pathetic.
 
I am reducing Brault's FTS solar flux and intensity spectra taken at
Kitt Peak for .3 to 5 $\mu$m.
I am trying to compute the telluric spectrum and ratio it out to determine
the flux above the atmosphere but that will not work for regions of very strong
absorption.  Once that is done the residual flux spectra can be normalized to low
resolution calibrations to determine the irradiance spectrum.  The missing pieces
will have to be filled in by computation.  Spectra available in the ultraviolet
are much lower resolution, much lower signal-to-noise, and are central
intensity or limb intensity, not flux.  The details of the available solar
atlases can be found in two review papers, Kurucz (1991; 1995a).
\ls
\ls
\noindent
{\bf We do not know how to determine abundances}
 
\noindent
{\bf We do not know the abundances of the sun or any other star}
 
\ls
One of the curiosities of astronomy is the quantity [Fe].  It is the logarithmic
abundance of Fe in a galaxy, cluster, star, whatever, relative to the solar
abundance of Fe.  What makes it peculiar is that we do not yet know the solar
abundance of Fe and our guesses change every year.  The abundance has varied
by a factor of ten since I was a student.  Therefore [Fe] is meaningless unless
the solar Fe abundance is also given so that [Fe] can be corrected to the
current value of Fe.
 
For an example I use Grevesse and Sauval's (1999) solar Fe abundance
determination.  I am critical, but, regardless of my criticism,
I still use their abundances.  There are scores of other abundance analysis
papers, including some bearing my name, that I could criticize the same way.
 
Grevesse and Sauval included 65 Fe I ``lines" ranging in strength from 1.4
to 91.0 m\AA and 13 Fe II ``lines" ranging from 15.0 to 87.0 m\AA.  They found
an abundance log Fe/H + 12 = 7.50 $\pm$ 0.05.
 
Another curiosity of astronomy is that Grevesse and Sauval have decided a priori
that the solar Fe abundance must equal the meteoritic abundance of 7.50 and that a
determination is good if it produces that answer.  If the solar abundance is not
meteoritic, how could they ever determine it?
 
There are many ``problems" in the analysis.   First, almost all the errors are
systematic, not statistical.  Having many lines in no way decreases the error.
In fact, the use of a wide range of lines of varying strengths increases the
systematic errors.  Ideally a single weak line is all that is required to
get an accurate abundance.  Weak lines are relatively insensitive to the
damping treatment, to microturbulent velocity, and to the model structure.
The error is reduced simply by throwing out all lines greater than 30 m\AA.
That reduces the number of Fe I lines from 65 to 25 and of Fe II lines from
13 to 5.  As we discussed above, the microturbulent velocity varies with
depth but Grevesse and Sauval assume that it is constant.  This  problem is
minimized if all the lines are weak.
 
As we discussed above ``lines" do not exist.  The lines for which equivalent
widths are given are all parts of blended features.  As a minimum we have to
look at the spectrum of each feature and determine how much of the feature
in the ``line" under investigation and how much is blending.  Rigorously one
should do spectrum synthesis of the whole feature.  We have  solar central
intensity spectra and spectrum synthesis programs.  For the sun we have
the advantage of intensity spectra without rotational broadening.  In the flux
spectrum of the sun and of other stars there is more blending.  The
signal-to-noise of the spectra is several thousand and the continuum level
can be determined to on the order of 0.1 per cent so the errors from the
spectrum are small.  With higher signal-to-noise more detail would be visible
and the blending would be better understood.  Most of the features cannot be
computed well with the current line data.
None of the features can be computed well without adjusting the line data.
Even if the line data were perfect, the wavelengths would still have to be
adjusted because of wavelength shifts from convective motions.
 
Fe has 4 isotopes.  The isotopic splitting has not been determined for the
lines in the abundance analysis.  For weak lines it does not affect the total
equivalent width but it does affect the perception of blends.
 
It is possible to have undetectable blends.  There are many Fe I lines with
the same wavelengths, including some in this analysis, and many lines of other
elements.  We hope that these blends are
very weak.  The systematic error always makes the observed line stronger than it is
in reality so they produce an abundance overestimate.
 
There are systematic errors and random errors in the gf values.  With a small
number of weak lines on the linear part of the curve of growth it is easy to
correct the abundances when the gf values are improved in the future.
 
We are left with 3 relatively safe lines of Fe I and 1 relatively safe line of
Fe II.  These have the least uncertainty in determining the blending by my
estimation.  Grevesse and Sauval found abundances of 7.455, 7.453, and 7.470
for the Fe I lines and 7.457 for the Fe II line.

\ls
\ls
\noindent
{\bf We do not have good atomic and molecular data}
 
\noindent
{\bf One half the lines in the solar spectrum are not identified}
\ls
It is imperative that laboratory spectrum analyses be improved and extended,
and that NASA and ESA pay for it.  Some of the analyses currently in use
date from the 1930s and produce line positions uncertain by 0.01 or 0.02 \AA.
New analyses with FTS spectra produce many more energy levels  and one or two
orders of magnitude better wavelengths.  One analysis can affect thousands of
features in a stellar spectrum.  Also the new data are of such high quality
that for some lines the hyperfine or isotopic splitting can be directly measured.
Using Pickering (1996) and Pickering and Thorne (1996) I am now able to compute
Co I hyperfine transitions and to reproduce the flag patterns and peculiar
shapes of Co features in the solar spectrum.  Using Litzen, Brault, and Thorne
(1993) I am now able to compute the five isotopic transitions for Ni I
and to reproduce the Ni features in the solar spectrum.  These new analyses
also serve as the basis for new semiempirical calculations than can predict
the gf values and the lines that have not yet been observed in the lab but
that matter in stars.  I have begun to compute new line lists for all the
elements and I will make them available on my web site, kurucz.harvard.edu.
\ls
\ls
\noindent
{\bf We should get our own house in order before worrying about the neighbors.}
\ls
\ls
\noindent
Ayres, T.R. and Testerman, L. 1981, ApJ 245, 1124-1140.
 
\noindent
Bikmaev, I. 1994, personal communication.
 
\noindent
Blackwell, D.E., Lynas-Gray, A.E., and Smith, G. 1995, A\&A, 296, 217.
 
\noindent
Chan, K.L., Nordlund, \AA, Steffen, M., Stein, R.F. 1991, {\it The Solar Interior and
 
Atmosphere}, A.N. Cox, W.C. Livingston, and M. Matthews, eds. (Tucson:
 
U. of Arizona Press) 223-274.
 
\noindent
Grevesse, N. and Sauval, A.J. 1999, A\&A 347, 348-354.

\noindent
Kurucz, R.L. 1991, in {\it The Solar Interior and Atmosphere} A.N. Cox,
 
W.C. Livingston, and M. Matthews, eds., Tucson: U. of Arizona Press, 663-669.
 
\noindent
Kurucz, R.L. 1993, Physica Scripta, T47, 110-117,

\noindent
Kurucz, R.L. 1995a. in {\it Laboratory and Astronomical High Resolution Spectra}
 
ASP Conf. Series 81, (eds. A.J. Sauval, R. Blomme, and N. Grevesse)  17-31.
 
\noindent
Kurucz, R.L. 1995b. Ap.J.,452, 102-108.
 
\noindent
Kurucz, R.L. 1996. in ASP Conf. Series 108, {\it Model Atmospheres and Stellar
 
Spectra} (eds. S. Adelman, F. Kupka, and W.W. Weiss)  2-18.
 
\noindent
Kurucz, R.L, Furenlid, I., Brault, J., and Testerman, L. 1984. {\it Solar Flux Atlas from
 
296 to 1300 nm} (Sunspot, N.M.: National Solar Observatory)
 
\noindent
Litzen, U., Brault, J.W., and Thorne, A.P. 1993, Physica Scripta 47, 628-673.
 
\noindent
Nordlund, \AA. and Dravins, D. 1990, A\&A 228, 155.
 
\noindent
Pickering, J.C. 1996, ApJ Supp 107,  811-822.
 
\noindent
Pickering, J.C. and Thorne, A.P. 1996, ApJ Supp 107,  761-809.
 
\noindent
 
\noindent
Topka, K.P. and Title, A.M. 1991, in {\it The Solar Interior and Atmosphere}, A.N. Cox,
 
W.C. Livingston, and M. Matthews, eds. (Tucson: U. of Arizona Press) 727-747.

\bye